\documentclass[twocolumn,preprintnumbers,amsmath,amssymb]{revtex4-1}

\usepackage{graphicx}
\usepackage{dcolumn}
\usepackage{bm,epsfig}
\usepackage{epstopdf}

\begin{document}

\title{Enhanced hybridization sets the stage for electronic nematicity in CeRhIn$_{5}$}

\author{P. F. S. Rosa$^{1}$, S. M. Thomas$^{1}$, F. F. Balakirev$^{2}$, E. D. Bauer$^{1}$, R. M. Fernandes$^{3}$, J. D. Thompson$^{1}$, F. Ronning$^{1}$, and M. Jaime$^{2}$}
\affiliation{
$^{1}$  Los Alamos National Laboratory, Los Alamos, New Mexico 87545, U.S.A.\\
$^{2}$ National High Magnetic Field Laboratory, Los Alamos, New Mexico 87545, U.S.A.\\
$^{3}$ School of Physics and Astronomy, University of Minnesota, Minneapolis, Minnesota 55455, USA.}
\date{\today}

\begin{abstract}
High magnetic fields induce a pronounced in-plane electronic anisotropy in the tetragonal antiferromagnetic metal CeRhIn$_{5}$ at  $H^{*} \gtrsim 30$~T  for fields 
$\simeq 20^{\mathrm{o}}$ off the $c$-axis. Here we 
investigate the response of the underlying crystal lattice in magnetic fields to $45$~T via high-resolution dilatometry. 
Within the antiferromagnetic phase of CeRhIn$_{5}$, a finite magnetic field component in the tetragonal $ab$-plane
explicitly breaks the tetragonal ($C_{4}$) symmetry of the lattice well below $H^{*}$ revealing a finite nematic susceptibility at low fields.
 A modest magnetostriction anomaly, $dL/L = -1.8 \times 10^{-6}$, at $H^{*} = 31$~T hence presumably marks the 
crossover to a fluctuating nematic phase with large electronic nematic susceptibility. Magnetostriction quantum oscillations confirm a Fermi surface change at $H^*$ with the emergence of new orbits. By analyzing 
the field-induced
change in the crystal-field ground state, we conclude that the in-plane Ce $4f$ hybridization is enhanced at $H^*$, carrying the in-plane $f$-electron anisotropy to the Fermi surface. We argue that the nematic behavior observed in this 
prototypical heavy-fermion material is of electronic origin, and is driven by the hybridization between $4f$ and conduction electrons.

 \end{abstract}

\maketitle

For more than half a century, the investigation of rare-earth-based materials has provided predictive understanding in both fundamental and applied realms \cite{RE1}. From the quantum theory of magnetism to the 
design of 
magnets in hard-drives, these materials, synthesized with $4f$ open-shell elements, have stimulated a diverse set of scientific discoveries. Cerium-based materials are a 
particularly intriguing case because their $4f$ electron may hybridize with the sea of conduction electrons \cite{Ce}. The $f$-electron delocalization destabilizes the otherwise magnetically ordered ground 
state and novel quantum phenomena may arise at the Fermi surface (FS).

Heavy electron masses, unconventional superconductivity, and non-Fermi-liquid behavior are a few known examples of emergent phenomena in Ce-based
 materials \cite{NFL}. More recently,
the discovery of a large electronic in-plane anisotropy induced by high magnetic fields in tetragonal CeRhIn$_{5}$ reveals the possibility of yet another 
novel state of 
matter, the so-called XY nematic \cite{Nematic-2017}. In an electronic nematic phase, the symmetry of the electronic system is lowered compared to that 
of the underlying lattice, in analogy to the directional alignment in 
nematic liquid crystals with continuous translational symmetry \cite{Nematic-Review1,Nematic-Review2}. Above an out-of-plane critical field of $H^* \approx 30$~T, but within the
antiferromagnetic (AFM) phase, electrical resistivity
measurements reveal electronic nematicity in CeRhIn$_{5}$. The small in-plane field component necessary to break the rotational symmetry of the electronic structure suggests a remarkably large nematic 
susceptibility. Moreover, the small 
magnitude of the magnetostriction anomaly at $H^{*}$, along with a similar response in the $B_{1g}$ and $B_{2g}$ channels, indicates that this phase is not strongly pinned to the lattice, placing CeRhIn$_{5}$ as a rare
XY-nematic candidate. At $H^*$, torque magnetometry measurements also identify a FS reconstruction and point to a larger FS in the 
nematic phase \cite{PNAS, PhilMag}.

At zero field and zero pressure, CeRhIn$_{5}$ undergoes a phase transition to a helix AFM order at $T_{N}=3.8$~K with 
ordering wavevector $\mathbf{Q}_{1}=(0.5,0.5,0.297)$ \cite{BaoCeRhIn5, Fobes1, Fobes2}.
 Pressurizing CeRhIn$_{5}$ tunes $T_{N}$ toward a quantum-critical 
point (QCP) at $P_{c2} = 2.3$~GPa and induces unconventional 
superconductivity around it \cite{TusonNature,Hegger, mito}. At $P_{c2}$ the effective electron mass diverges and the 
FS changes abruptly \cite{shishido}. 
Applied pressure is a clean tuning parameter known to increase the hybridization between $4f$ and conduction electrons, and, as a consequence, the 
$T$-$P$ phase diagram of CeRhIn$_{5}$ can be qualitatively understood in terms of the strength of the (Kondo) coupling between $4f$ and 
conduction electrons \cite{Leticie}. Remarkably, 
magnetic fields also destabilize the AFM order in CeRhIn$_{5}$ towards a QCP at $H_{c} \approx 50$~T and a FS reconstruction is observed 
at $H^* \approx 30$~T \cite{PNAS}. Magnetic fields, however, are 
symmetry breaking and expected to localize $4f$ electrons. Understanding why the Kondo coupling is robust in high fields is a first step towards 
unveiling the nature of the nematic phase. An additional relevant question is whether the $H^*$ boundary is a true (first-order) phase transition or a crossover.

To answer these questions, probes other than electrical resistivity are imperative. Various thermodynamic probes such as torque 
magnetometry, magnetic 
susceptibility, and specific heat, however, fail to observe an anomaly at $H^*$, although their response is affected above it.
Apart from the FS reconstruction seen by magnetometry, specific heat measurements in pulsed fields show a clear difference in the shape of the peak at $T_{N}$ across 
$H^*$ \cite{arxiv}. In this Letter, we use 
high-resolution dilatometry to probe the response of the underlying crystal lattice to magnetic fields. The finite coupling between the nematic phase and the lattice 
yields an anomaly at $H^*$ that can be tracked to high temperatures and vanishes above the AFM boundary.  A finite in-plane field component explicitly breaks the tetragonal ($C_{4}$) symmetry of the lattice 
revealing a finite nematic susceptibility at low fields. The small magnetostriction anomaly at $H^{*} = 31$~T hence marks the 
crossover to a fluctuating nematic phase with large nematic susceptibility. This crossover occurs concomitantly to a Fermi surface change at $H^*$ and an 
enhancement in the Ce $4f$ hybridization with the in-plane In(1) conduction electrons, suggesting that the nematic
phase stems from the $4f$ degrees of freedom and their anisotropy which is translated to the Fermi surface via hybridization. The Fermi surface change is confirmed by  unprecedented quantum oscillations in the 
magnetostriction of CeRhIn$_{5}$ that reveal the emergence of new orbits. 

Single crystals of CeRhIn$_{5}$ were grown by the Indium-flux technique. The crystallographic structure and orientation were verified by X-ray powder diffraction and 
Laue diffraction at $300$~K, respectively.  
Low-field magnetostriction measurements were performed using the standard capacitance dilatometry technique in a Quantum Design PPMS above $2$~K \cite{George}. The longitudinal and transverse magnetostriction of 
CeRhIn$_{5}$ was obtained as a function of DC fields to $9$~T applied parallel 
 to the $a$-axis with a resolution in $dL/L$ of $10^{-8}$.
High-field magnetostriction measurements were performed using optical Fiber Bragg Grating (FBG) sensors in a hybrid magnet at $^3$He temperatures \cite{Marcelo}. 
The $c$-axis magnetostriction, $dL_{c}/L_{c}$, was obtained as a function of DC fields to $45$~T applied $\simeq 20^{\mathrm{o}}$ off the $c$-axis. The data was obtained with a swept wavelength laser Micron Optics interrogator, 
yielding a resolution of $3\times10^{-8}$.

The first point we will address is how the system responds to an in-plane magnetic field that explicitly breaks tetragonal symmetry. 
Figure 1 shows the anisotropic magnetostriction of CeRhIn$_{5}$ for fields applied along the $a$-axis. When $dL||H||a$, the longitudinal magnetostriction is negative and 
displays a sharp contraction at $2.2$~T, signaling the transition from a helix state (AF1) to a commensurate collinear square-wave state (AF3) with ordering vector $\mathbf{Q_{3}}=(1/2,1/2,1/4)$ \cite{Fobes2}.  The transverse magnetostriction ($dL||b$), however, is 
positive 
 and displays a sharp expansion at $2.2$~T. In a conventional material, compression along the $a$-axis produces an expansion along the perpendicular axes (i.e. $b$- and $c$-axes), and the ratio of the 
 perpendicular 
 strains is known as the Poisson's ratio, $\nu_{ij}$ \cite{Poisson}. In particular, $\nu_{xy}$ gives the in-plane strain response. The calculated Poisson's ratio in this scenario is 
 $\nu_{xy} \equiv s_{xy}/s_{xx}$ = 0.2, where $s_{ij}$
 are the elastic compliances at $10$~K obtained from the inverted elastic modulus tensor \cite{Kumar}. 
The experimental ratio between the transverse and longitudinal magnetostriction response, however, is not only larger than the calculated one, but also field- and temperature-dependent. This difference suggests that there are other degrees of freedom contributing to the observed response.  

To describe these observations and model the magneto-elastic coupling in this material, we write the magneto-elastic free energy of a tetragonal system 
as $F~=~-~\lambda \delta (H_{a}^{2} - H_{b}^{2}) + (\alpha/2) \delta^{2}$, where
$\delta = (a-b)/a$ is the orthorhombic distortion, $\lambda$ is a 
coupling constant, and $\alpha$ is the elastic constant renormalized by nematic fluctuations acting as an indirect measure of the nematic susceptibility. As a consequence, one can estimate the nematic susceptibility, $\chi_{\mathrm{nem}}$,
from the data via $\chi_{\mathrm{nem}} \propto \partial \delta/\partial H_{a}^{2}$, where $\delta = dL_{a}/L_{a} - dL_{b}/L_{b}$. Our results 
show
 that $\gamma \chi_{\mathrm{nem}} = -2 \times 10^{-9}$ in the AF1 phase and $\gamma \chi_{\mathrm{nem}} = -6 \times 10^{-10}$ in the AF3 phase. 
The nematic susceptibility is finite, but curiously smaller in AF3. This result suggests that $\chi_{\mathrm{nem}}$ is small in the low 
field phases and is not enhanced as a function of magnetic fields. In fact, the nematic response in the electronic degrees of freedom, i.e., the in-plane
 resistivity anisotropy, is vanishingly small at low fields and increases sharply at $H^{*}$ (inset of Fig.~1). We note that the nematic order parameter couples to any quantity that breaks tetragonal symmetry. As a consequence, nematic fluctuations $-$ and hence the corresponding nematic susceptibility $-$ can be indirectly probed in different quantities, such as the elastic constant $\alpha$ or the rate of change of the anisotropic resistivity,
 with different coupling strengths.

\begin{figure}[!ht]
\begin{center}
\vspace{-0.2cm}
\hspace{-0.35cm}
\includegraphics[width=0.8\columnwidth]{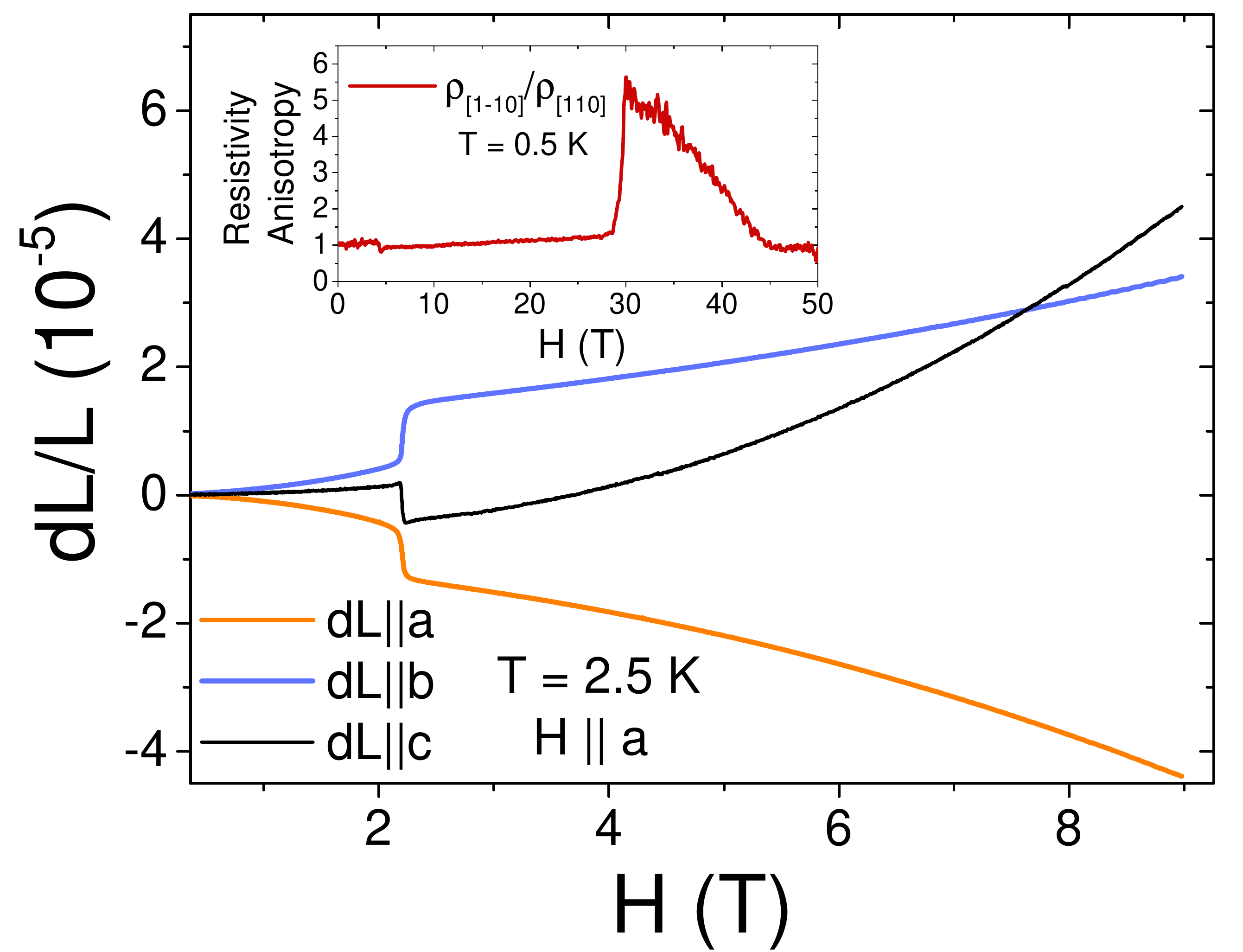}
\vspace{-0.65cm}
\end{center}
\caption{Magnetostriction of CeRhIn$_{5}$ along the $a$-axis, $dL_{a}/L_{a}$, at 2.5~K for fields applied in the tetragonal $ab$-plane. Inset shows the in-plane anisotropy in 
electrical resistivity \cite{Nematic-2017}.}
\label{fig:Fig1}
\vspace{-0.2cm}
\end{figure}

Next, we turn our attention to the lattice response at fields applied 20$^{\mathrm{o}}$ off the $c$-axis at which the in-plane resistivity anisotropy 
 is pronounced. Figure~2 shows the 
$c$-axis magnetostriction, $dL_{c}/L_{c}$, to $45$~T at the base temperature 
of our measurements, $350$~mK.
At low fields, a broad feature is observed at $H_{MM}=7.6~$T reminiscent of the AF1-AF3 transition discussed above (Fig.~1).  Applying the known $1/\mathrm{sin}(\theta)$ dependence of this  
transition, we obtain that the 
actual angle between the $c$-axis and the magnetic field is $\theta =17^{\mathrm{o}}$. At higher fields, a small anomaly is buried in the background and the inset of Fig.~2 shows $dL_{c}/L_{c}$ after the subtraction of 
a 4th order polynomial fit obtained in the range $15 < H < 29$~T. The subtracted data show a small lattice contraction at $H^{*}=31$~T, $dL/L = -1.8 \times 10^{-6}$, in agreement with the upper limit of $-2 \times 10^{-6}$ obtained recently in pulsed fields \cite{Nematic-2017}. Remarkably, there is no noticeable hysteresis in 
our data,
which does not match the hysteretic resistivity obtained in thin microstructured samples, but agrees with virtually hysteresis-free resistivity curves obtained in larger samples \cite{NatComm}.
The lack of hysteresis along with a broader transition at $H^{*}$ in bulk samples suggests that the nematic transition might actually be a crossover to a regime with large nematic susceptibility, which in turn is highly sensitive to strain.
As elaborated in Ref. \cite{NatComm}, the 
microstructured samples are coupled to the substrate, and the strain relaxation might be different in thinner samples. As a result, smaller samples may take longer to relax to equilibrium and hence 
may display hysteresis.

\begin{figure}[!ht]
\begin{center}
\hspace{-0.35cm}
\includegraphics[width=0.8\columnwidth]{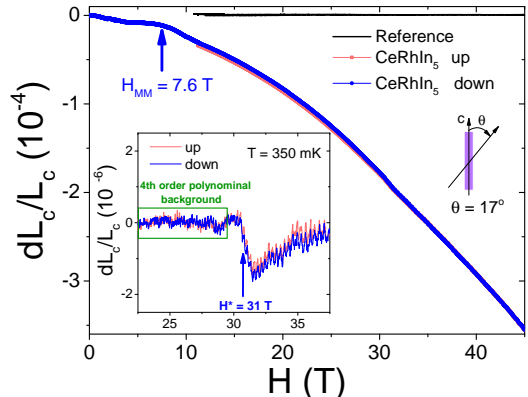}
\vspace{-0.65cm}
\end{center}
\caption{Magnetostriction of CeRhIn$_{5}$ along the $c$-axis, $dL_{c}/L_{c}$, at $T= 350$~mK for fields applied 20$^{\mathrm{o}}$ off the $c$-axis. Inset shows the data after a background subtraction.} 
\label{fig:Fig2}
\vspace{-0.25cm}
\end{figure}

The inset of Fig.~2 also shows sizable quantum oscillations at $H \gtrsim 27$~T. Although the presence of quantum oscillations (QO) in magnetostriction is a phenomenon known since the 60s \cite{MS-first}, this is the first time 
magnetostriction QO are reported in CeRhIn$_{5}$, likely due to the fact that high-quality single crystals, high-resolution 
dilatometers, and low-noise environments (i.e. DC fields) are required. 
The amplitude of oscillations along the $c$-axis can be written as $-MH (\partial \mathrm{ln} A/ \partial \sigma_{c}$), where $M$ is  the amplitude of 
oscillations in magnetization,  $A$ is the extremal cross-sectional area of the FS perpendicular to the applied magnetic field, and $\sigma_{c}$ is the stress along the $c$-axis. 
Therefore, magnetostriction QO
provide information on the Fermi surface of materials, as do magnetization QO (i.e. dHvA), but the additional term $\partial \mathrm{ln} A/ \partial \sigma_{c}$ enhances the amplitude of the 
stress-sensitive orbits. 
Here we will focus on the region  $H > H^*$ because the QO amplitudes are larger above $H^*$, and our noise floor is higher in comparison to dHvA measurements \cite{PNAS}. 

Figure 3a shows the magnetostriction oscillations in inverse field and Figure 3b shows the 
corresponding FFT spectra at $350$~mK. We observe four frequencies 
($\alpha_{3}=3.7$~kT, $\alpha'_{2}=4.7$~kT, $\alpha'_{1}=5.7$~kT, and $\beta_{2}=6.2$~kT) which agree well with the frequencies obtained
 via dHvA for $30 <H < 45$~T at $330$~mK and fields along the $c$-axis
($\alpha_{3}=3.7$~kT, $\alpha'_{2} = 5$~kT, $\alpha'_{1}=5.7$~kT, and $\beta_{2}=6.3$~kT) \cite{PhilMag}. 
We note that dHvA frequencies marked as $\alpha'_{2}$ and $\alpha'_{1}$ are observed 
only above $H^*$.
By comparing the amplitude of the orbits given by the different techniques in similar conditions, one can obtain
information on the strain dependence of the orbits. For instance, the $\beta_{2}$ amplitude is reduced in our data as compared to dHvA results whereas 
the $\alpha'_{2}$ is relatively 
enhanced. These results 
suggest that $\alpha'_{2}$ orbit is a more strain-sensitive orbit than $\beta_{2}$. We note, however, that this comparison should be taken with caution 
because magnetostriction and dHvA measurements were not performed on the same 
crystal and there may be differences in the FFT analysis performed in different experiments.

\begin{figure}[!ht]
\begin{center}
\hspace{-0.25cm}
\includegraphics[width=0.85\columnwidth,keepaspectratio]{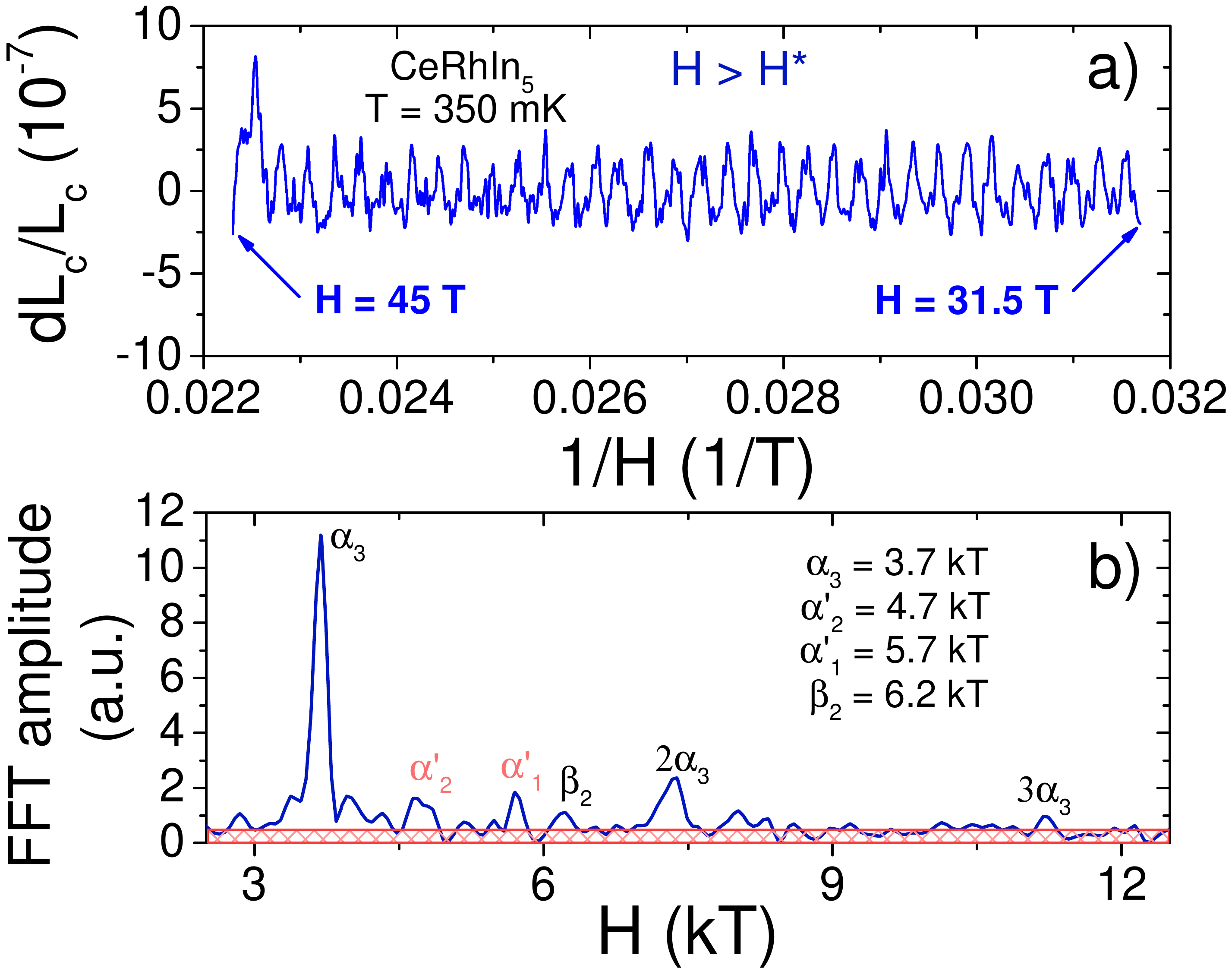}
\vspace{-0.5cm}
\end{center}
\caption{a) Magnetostriction of CeRhIn$_{5}$ at $350$~mK as a function of inverse field for fields applied $ \simeq 20^{\mathrm{o}}$ off the $c$-axis. For the sake of clarity, a high-pass filter 
was used to remove low-frequency oscillations that likely originate from the background difference below and above $H^*$. b) FFT spectra in the region $31.5 < H < 45$~T. The dashed area is an estimate
of the noise floor.}
\label{fig:Fig3}
\vspace{-0.3cm}
\end{figure}

\begin{figure}[!ht]
\vspace{-0.2cm}
\begin{center}
\hspace{-0.05cm}
\includegraphics[width=0.85\columnwidth,keepaspectratio]{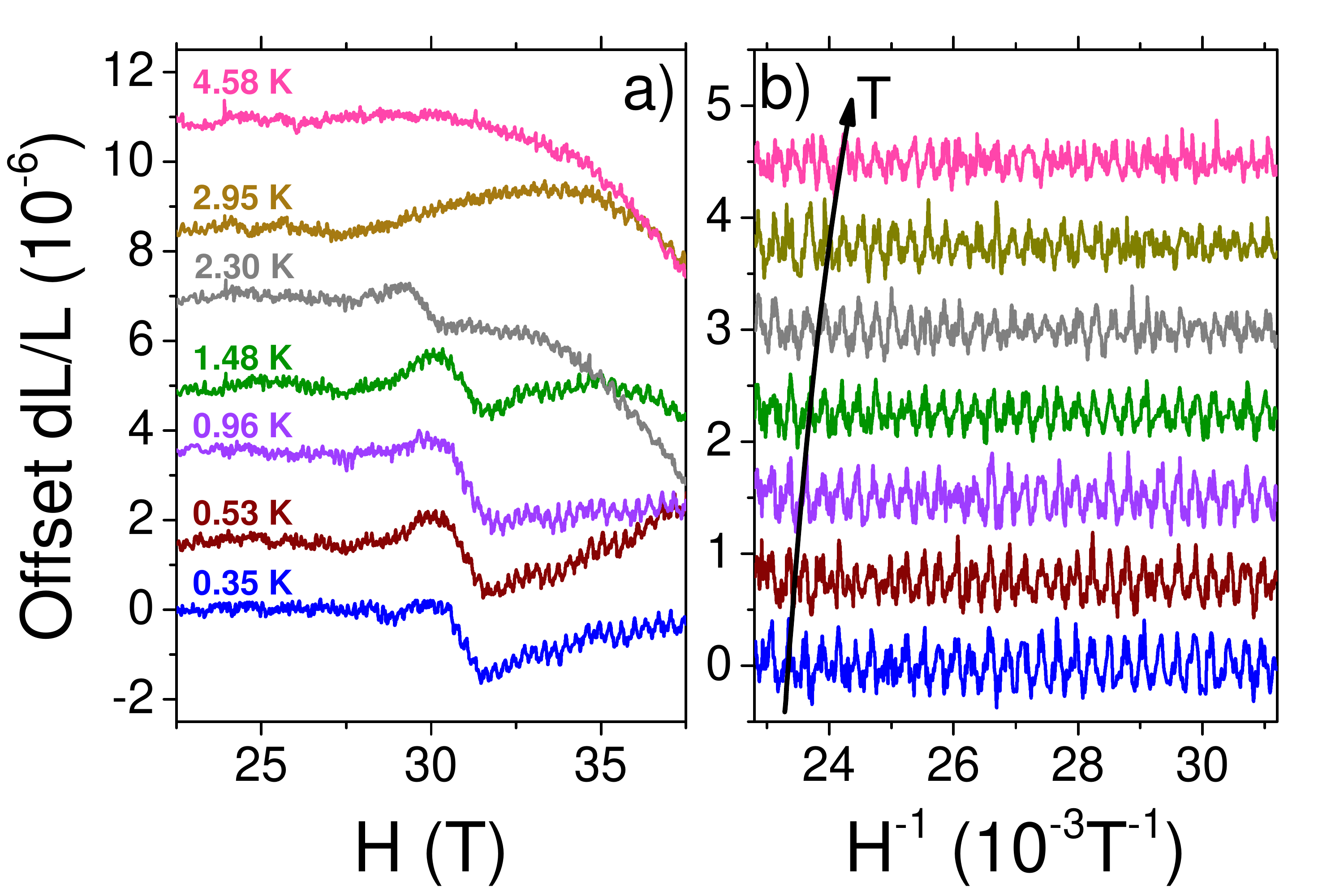}
\vspace{-0.5cm}
\end{center}
\caption{a) Temperature dependence of the magnetostriction of CeRhIn$_{5}$ for fields applied $\simeq 20^{\mathrm{o}}$ off the $c$-axis. An offset is added for the sake of clarity. b) High-field quantum oscillations in inverse 
field obtained from the data shown in panel a) after a high-pass filter is employed. The color scheme is the same as used in (a).}
\label{fig:Fig4}
\vspace{-0.5cm}
\end{figure}

Figure 4a shows the $c$-axis magnetostriction at various temperatures for fields applied $\simeq 20^{\mathrm{o}}$ off the $c$-axis. As the temperature increases, $H^*$ remains fairly 
constant, but deviates to slightly lower fields at $2.3$~K before vanishing at higher temperatures. We 
note that the anomaly at $H^*$ detected in electrical resistance measurements becomes unobservable 
at $ T > 2.2$~K, which could be a consequence of resolution limitations. By tracking $H^*$ to higher temperatures, we are able to affirm that the 
nematic boundary most likely intercepts the AFM boundary and disappears above $T_{N}$, as shown in the $H-T$ phase diagram to be discussed below (Fig.~5).

Figure~4b shows the high-field quantum oscillations in magnetostriction with increasing temperature. Although we are unable to reliably extract the 
$T$-dependence of the orbits $\alpha'_{2}$, $\alpha'_{1}$, and $\beta_{2}$ due to their small amplitudes, $\alpha_{3}$ is significantly more intense and can be tracked to 
higher temperatures. Remarkably, the overall amplitude of the oscillations does not display the expected behavior for conventional metals within the 
Lifschitz-Kosevich (LK) formalism. 
According to the LK formula, the amplitude of the oscillations in a particular field range increases with decreasing temperature 
as $\propto m^*T/\mathrm{sinh}(Cm^*T)$, where $C$ is a constant and $m^*$ the effective mass \cite{QO}.
Although the amplitude of the oscillations do decrease when comparing the temperature extremes (4.58~K and 0.35~K), there is no clear trend below $1$~K. This is in agreement with dHvA measurements,  
both in pulsed and DC fields, that observe a decrease in the amplitude of the $\alpha_{3}$ orbit below about $1$~K  \cite{Cornelius, arxiv}. The interpretation of this 
anomalous behavior, however, is not settled.  On one hand, the early
report in pulsed fields attributes this decrease to the formation of spin-density-wave order \cite{Cornelius}. On the other hand, more recent dHvA results in DC fields 
support a spin-dependent mass enhancement of the FS. In the latter, the QO amplitude is well described  by a spin-dependent LK formula, suggesting 
a spin-split FS, as observed previously in CeCoIn$_{5}$ \cite{arxiv, Co-LK}. In fact, we observe a beating pattern at 
frequencies close to $\alpha_{3}$, which could be taken as indicative of two Fermi pockets close together, one spin-up and one spin-down.

Finally, we discuss the implications of our results for the nature of the nematic phase. Figure~5 displays the $H$-$T$ phase diagram of CeRhIn$_{5}$ with 
a compilation of recent high-field data.  Because this phase diagram is constructed in the presence of a symmetry-breaking magnetic field,
the phase boundary at $H^*$ may be a crossover, instead of a true phase transition. In fact, removing the magnetic field does not seem to result in a residual resistivity anisotropy \cite{Nematic-2017}, 
which is indicative of a large nematic susceptibility, but no long-range nematic order.

If not long-range order, what changes at $H^*$ that causes a large nematic susceptibility and why is the Kondo coupling robust in high magnetic fields?
We recall that recent dHvA measurements reveal that the FS changes at $H^*$ pointing to a larger FS in the nematic phase \cite{PNAS}. 
This indicates that the $4f$ electrons are being incorporated to the FS at $H^*$ and are becoming more itinerant. 
Therefore, the main tuning parameter here must be the strength of the 
$4f$-conduction electron hybridization. The key point therefore is that the hybridization depends on the 
Ce $4f$ ground-state wavefunction, which in turn is given by the crystal-field parameters in this particular
tetragonal structure. Thus, the answer to our question lies in the field dependence of the wavefunctions determined by the 
crystalline electric field (CEF) and their anisotropic hybridization. For CeRhIn$_{5}$, the low-energy CEF levels are given by \cite{severing1}:

\vspace{-0.15cm}
\begin{equation}
\begin{split}
|0\rangle = \Gamma_{7}^{2} & = \alpha |\pm 5/2\rangle - \beta |\mp 3/2\rangle,\: \alpha = 0.62, \\
|1\rangle = \Gamma_{7}^{1} & = \beta |\pm 5/2\rangle + \alpha |\mp 3/2\rangle, \: \beta = 0.78
\end{split}
\end{equation} 
\noindent where $|0\rangle$ is the ground state, $|1\rangle$ is the first-excited state at $7$~meV, and $\alpha^2$ determines the out-of-plane anisotropy. 
The pure $|5/2\rangle$ orbital is donut-shaped and hence higher $\alpha^2$ corresponds to a more oblate $4f$ orbital confined to the $ab$-plane.
Linear-polarized soft-x-ray absorption experiments reveal that the ground-state doublet changes from flatter (i.e. larger $|5/2\rangle$) orbitals in CeRhIn$_{5}$ to 
orbitals that are more extended along the $c$-axis in CeIrIn$_{5}$ and CeCoIn$_{5}$ (i.e. smaller $|5/2\rangle$) \cite{severing2}.
 The prolate orbitals hybridize more strongly with 
out-of-plane In(2) electrons and, as a consequence, lead to superconducting ground states in CeIrIn$_{5}$ and CeCoIn$_{5}$. 

\begin{figure}[!ht]
\vspace{-0.15cm}
\begin{center}
\hspace{-0.75cm}
\includegraphics[width=0.85\columnwidth,keepaspectratio]{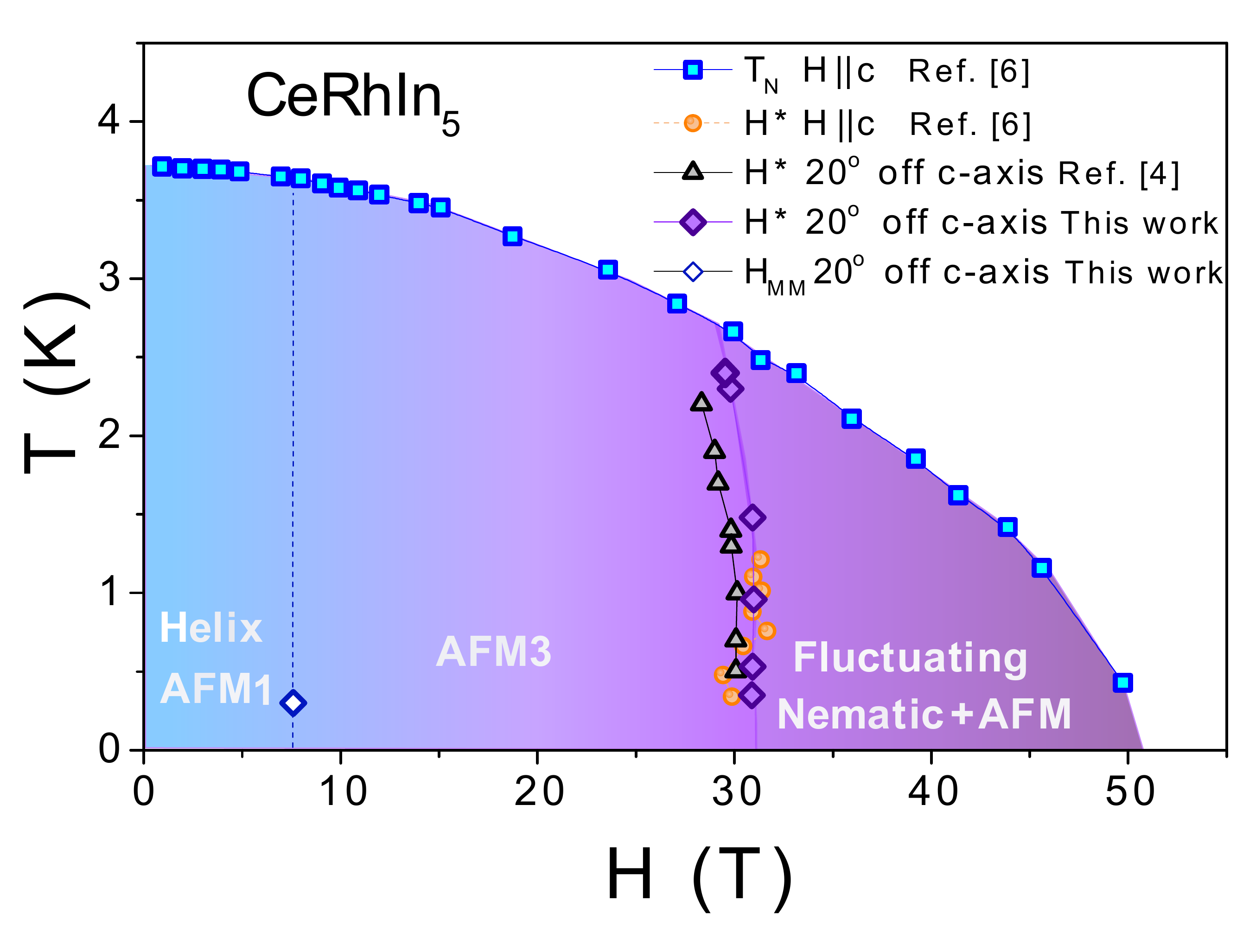}
\end{center}
\vspace{-0.75cm}
\caption{$H-T$ phase diagram of CeRhIn$_{5}$. Data from Ref. [6] were obtained by specific heat measurements and Hall resistivity. Data from Ref. [4] 
were obtained by electrical resistivity measurements.}
\vspace{-0.25cm}
\label{fig:Fig5}
\end{figure}

Now we can turn our attention to the field dependence of the orbitals in Eq. (1). Magnetic fields will, by the Zeeman effect, split the $\Gamma_{7}$ doublets and, 
therefore, promote mixing between the ground-state, $\Gamma_{7}^{2}$,
and the first-excited state, $\Gamma_{7}^{1}$. Interestingly, the $|5/2\rangle$ contribution in the first excited state ($\beta = 0.78$) is larger than that of
the original ground state ($\alpha = 0.62$), implying that 
the new, field-induced ground state wavefunction will become more confined to the basal plane.
Because of its modified shape, the new ground state displays an enhanced hybridization with the in-plane In(1) electrons, similar to what happens with Sn-doped
Ce$M$In$_{5}$ ($M=$Co, Rh) \cite{severing3}. We note that this hybridization increase
with In(1) electrons is fundamentally different from the hybridization with In(2) electrons observed in CeIrIn$_{5}$, CeCoIn$_{5}$, and likely CeRhIn$_{5}$
under pressure. This difference explains not only why the FS increases at $H^*$ with the field-induced incorporation of $4f$ electrons, but it may also be the reason
CeRhIn$_{5}$ is not superconducting as a function of field or Sn doping. Additionally, a similar crystal-field scenario has been used to explain recent nuclear magnetic resonance (NMR) experiments in high 
fields \cite{Urbano}. As a local probe, NMR is able to rule out a change in the magnetic structure at $H^*$.
Nevertheless, the fact that $H^*$ develops only inside the magnetically ordered state along with the correlation between enhanced hybridization and enhanced nematic susceptibility suggests that the latter stems from the $f$-electron degrees of freedom. Whether a consequence of the frustration and the field-induced magnetic anisotropy in the spin degrees of freedom known to exist in CeRhIn$_{5}$ \cite{Fobes2,Pinaki} or a consequence of the hybridization gap remains an open question.

In summary, we performed high-resolution magnetostriction measurements in CeRhIn$_{5}$ in DC fields to $45$~T. At low fields, a finite in-plane field 
explicitly breaks the tetragonal ($C_{4}$) symmetry of the underlying lattice and reveals a small nematic susceptibiliity. 
 At high fields, a small anomaly ($dL/L = - 1.8 \times 10^{-6}$) in the
$c$-axis magnetostriction marks the onset of the nematic response at $H^{*} = 31$~T in CeRhIn$_{5}$ for fields $\simeq 20^{\mathrm{o}}$ off the $c$-axis. The obtained $H-T$ phase diagram hosts a crossover line at $H^*$ to a fluctuating 
nematic phase with high nematic susceptibility. This crossover occurs concomitantly to an 
enhancement in the Ce $4f$ hybridization with the in-plane In(1) conduction electrons, which explains why the Kondo coupling is robust in high fields. Our results also suggest that the nematic
phase stems from the $4f$ degrees of freedom and their anisotropy translated to the Fermi surface via hybridization.
Therefore, the nematic behavior observed here in a prototypical heavy-fermion material is of electronic origin, and is driven 
by hybridization.

\begin{acknowledgments}
We would like to acknowledge constructive discussions with P. G. Pagliuso, R. R. Urbano, L. Jiao, A. Severing, and E. Miranda. 
P.~F.~S.~R. acknowledges support from the Laboratory Directed Research and Development program of Los Alamos National Laboratory under project number 
20180618ECR.  Sample synthesis was supported by the U.S. Department of Energy, Office of 
Basic Energy Sciences, Division of Materials Science and Engineering. M. J. acknowledges support from the Institute for Materials Science, LANL. 
A portion of this work was performed at the National High Magnetic Field Laboratory, which is 
supported by National Science Foundation Cooperative Agreement No. DMR-1157490 and the State of Florida. We thank J. B. Betts at the pulsed facility, and J. Billings and T. Murphy at the DC facility for their 
technical support.  Theory work (RMF) was supported by the Office of Basic Energy Sciences, U.S. Department of Energy, under award DE-SC0012336.

\end{acknowledgments}

\bibliography{basename of .bib file}

\end{document}